\documentclass[draft]{agujournal2019}
\usepackage{url} 
\usepackage{lineno}
\usepackage[inline]{trackchanges} 
\usepackage{soul}
\draftfalse

\journalname{Geophysical Research Letters}

\begin{document}

%
%


\title{Field evidence for the initiation of isolated aeolian sand patches}

%
%




\authors{P. Delorme \affil{1,2}, J. M. Nield \affil{1}, G. F. S. Wiggs\affil{3}, M. C. Baddock\affil{4}, N. R. Bristow\affil{5}, J.L. Best\affil{6}, K. T. Christensen \affil{7}, and P. Claudin\affil{8}  }


\affiliation{1}{School of Geography and Environmental Science, University of Southampton, Southampton, UK}
\affiliation{2}{now at: Energy and Environment Institute, University of Hull, Hull, UK}
\affiliation{3}{School of Geography and the Environment, University of Oxford, Oxford, UK}
\affiliation{4}{Geography and Environment, Loughborough University, Loughborough, UK }
\affiliation{5}{Mechanical Engineering, St Anthony Falls Laboratory, University of Minnesota, Minneapolis, USA}
\affiliation{6}{Departments of Geology, Geography and GIS, Mechanical Science and Engineering and Ven Te Chow Hydrosystems Laboratory, University of Illinois at Urbana-Champaign, USA}
\affiliation{7}{Departments of Mechanical, Materials and Aerospace Engineering and Civil, Architectural and Environmental Engineering, Illinois Institute of Technology, USA}
\affiliation{8}{Physique et Mécanique des Milieux Hétérogènes, CNRS, ESPCI Paris, PSL Research University, Université Paris Cité, Sorbonne Université, Paris, France}




\correspondingauthor{Pauline Delorme}{p.m.delorme@hull.ac.uk}




\begin{keypoints}
\item Sand patches can emerge on non-erodible surfaces.
\item Differing surface characteristics control particle behaviour. 
\item Field measurements demonstrate the key role of sand transport in bedform initiation.
\end{keypoints}

%
%

%
%


\begin{abstract}
Sand patches are one of the precursors to early-stage protodunes and occur widely in both desert and coastal aeolian environments. Here we show field evidence of a mechanism to explain the initiation of sand patches on non-erodible surfaces, such as desert gravels and moist beaches. Changes in sand transport dynamics, directly associated with the height of the saltation layer and variable transport law, observed at the boundary between non-erodible and erodible surfaces lead to sand deposition on the erodible surface. This explains how sand patches can form on surfaces with limited sand availability where linear stability of dune theory does not apply. This new mechanism is supported by field observations that evidence both the change in transport rate over different surfaces and in-situ patch formation that leads to modification of transport dynamics at the surface boundary.
\end{abstract}

\section*{Plain Language Summary}
Sand patches can be observed in various environments such as beaches and
gravel plains in deserts. Expected to be precursors of dunes when sediment
supply is limited, these bedforms are typically a few centimeters high and
present a reverse longitudinal elevation profile, with a sharp upwind edge and a smooth
downwind tail. Based on field measurements, we propose a
formation mechanism for these patches associated with the sensitive nature
of wind-blown sand transport to changing bed conditions: sand saltation is
reduced at the transition from a solid to an erodible surface, hence
favouring deposition on the patches. This allows us to explain their
typical meter-scale length as well as their asymmetric shapes.

%
%

%


%
%
%
%

\section{Introduction}
\begin{figure}
\includegraphics[width=\textwidth]{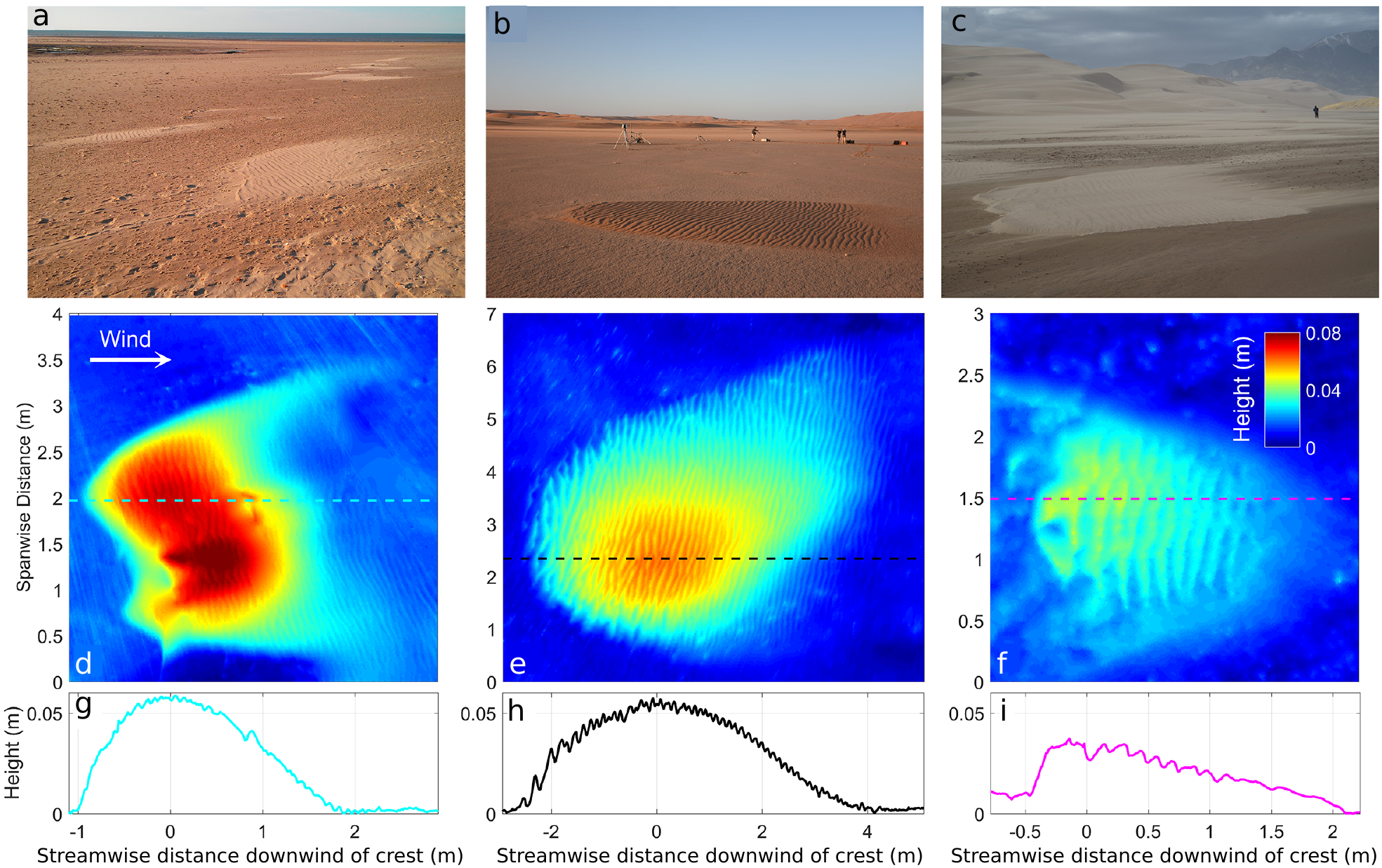}
\caption{Sand patches formed on different surfaces. Brancaster beach Norfolk, UK (a, d and g), Helga’s dunefield, Namib Desert, Namibia (b, e and h), and Medano Creek, Great Sand Dunes National Park, Colorado, USA (c, f and i). \label{Fig1}}
\end{figure}

Isolated low-angle sand patches are commonly observed in desert and coastal regions on non-erodible surfaces, such as gravel plains or moist beaches \cite<Figure \ref{Fig1}, e.g.>{lancaster1996field, hesp1997crescentic, nield2011surface}. These bedforms are typically several centimeters high, exhibit reverse longitudinal asymmetry compared to mature dunes, and can develop rapidly over several hours. Extensive research has explored the physical dynamics and morphology of mature desert sand dunes \cite{bagnold1937transport,bagnold1941physics,lancaster1982dunes,werner1990steady,andreotti2002selection,charru2013sand,courrech2015dune, wiggs2021}. We also have some evidence of the dynamics by which emerging dunes might grow into early-stage protodunes and more mature dune forms \cite{kocurek1992dune,nield2011aeolian, elbelrhiti2012initiation, hage2018determining, montreuil2020early}, where the subtle coupling of topography, wind flow, and sediment transport acts to reinforce their growth \cite{baddock2018early, delorme2020dune, gadal2020spatial, lu2021direct, bristow2022topographic}. However, our knowledge of the processes resulting in, and the relevant time and length scales associated with, the initial deposition of sand on a non-erodible surface remains incomplete and unquantified, although such processes possibly represent a fundamental stage in the origin of aeolian dunes.

There are two clear sets of processes by which aeolian dunes are thought to be established \cite{courrech2014two, courrech2015dune}. The first is associated with the hydrodynamic instability of an erodible granular flat bed with unlimited sand availability \cite{warren1979aeolian, andreotti2002selection, claudin2013field, charru2013sand}. This instability results from the combination of the response of wind stress to the modulated topographic profile, and the response of sand transport to the spatial variation in that wind stress \cite{charru2013sand}. The former drives the instability where the wind stress maximum is shifted upwind of a dune crest \cite{claudin2013field, lu2021direct}; the latter controls the emerging dune size with a relaxation process over a
(saturation) length, $L_{\mathrm{sat}}$ \cite{sauermann2001continuum, andreotti2010measurements, pahtz2013flux, selmani2018aeolian}. The resulting dune pattern consists of straight-crested bedforms growing in amplitude with an orientation controlled by the wind regime \cite{gadal2019incipient, delorme2020dune}. The second set of processes is associated with the growth of finger-like dunes developing across a non-erodible surface from isolated sand sources \cite{courrech2014two,rozier2019elongation, gadal2020periodicity}. In this case, the dunes, well separated by interdunes where sand is scarce, present a finger-like shape and grow in length in a direction between those of the dominant winds \cite{rozier2019elongation}. Experiments in wind tunnels have also highlighted the critical role of boundary conditions in determining saltation dynamics and sand transport rates \cite<e.g.>{ho2012particle, kamath2022scaling} and this offers a potential further means by which dunes may establish. These experiments have provided evidence for the existence of distinctly different transport rates on erodible and non-erodible or moist surfaces \cite{neuman1998wind, ho2011scaling}. Larger sediment fluxes on non-erodible beds have been interpreted as a consequence of a negligible feedback between the mobile grains on the flow. This is in contrast to the wind velocity `focal point' that exists when saltation takes place over an erodible granular bed where the saltating grains comprise a momentum sink on the overlying flow \cite{bagnold1937transport, ungar1987steady, creyssels2009saltating,duran2011aeolian, ho2014aeolian, valance2015physics}. 

Here, we propose a new mode for sand patch and protodune initiation associated with the sensitive nature of the transport law in response to changing bed conditions. We find that sand transport rates responding to non-erodible to erodible bed conditions can explain the emergence of isolated, meter-scale sand patches on gravelly interdune areas or moist beaches (Figure \ref{Fig1}). Our field data in support of this process, quantitatively capturing the emergence of a sand patch and the change in saltation this produces, allows us to propose a conceptual model for early-stage protodune growth from a flat bed.

\section{Methods}

Sediment transport measurements were undertaken in the Skeleton Coast National Park, Namibia on sand and gravel surfaces between the 13th and 15th September 2019. Here, wind speed was measured simultaneously on both surfaces using hotwire anemometers (DANTEC 54T35 probes) at a height of 0.085 m and a frequency of 0.1 Hz. Co-located sediment transport was measured via laser particle counters (Wenglor YH03PCT8, following the methods of \citeA{barchyn2014particle}), Sensit contact particle counters and modified Bagnold sand traps. Saltation height was measured, using a Leica P20 terrestrial laser scanner (TLS) following the methods of \citeA{nield2011application}, in a 1 m$^2$ area immediately upwind of the wind and sand transport instrument arrays, alternating between each of the gravel and sand sites.
Additional measurements were undertaken to quantify both saltation height and surface topographic change during the initial formation of a sand patch using Leica P20 and P50 TLS instruments placed downwind of an emerging patch at Great Sand Dunes National Park, Colorado, USA on the 15th April 2019. Details on the data processing methods can be found in the Supplementary Information.

\section{Evidence for Differing Sand Transport Processes on Surfaces with Different Erodibility}

\begin{figure}
\includegraphics[width=\textwidth]{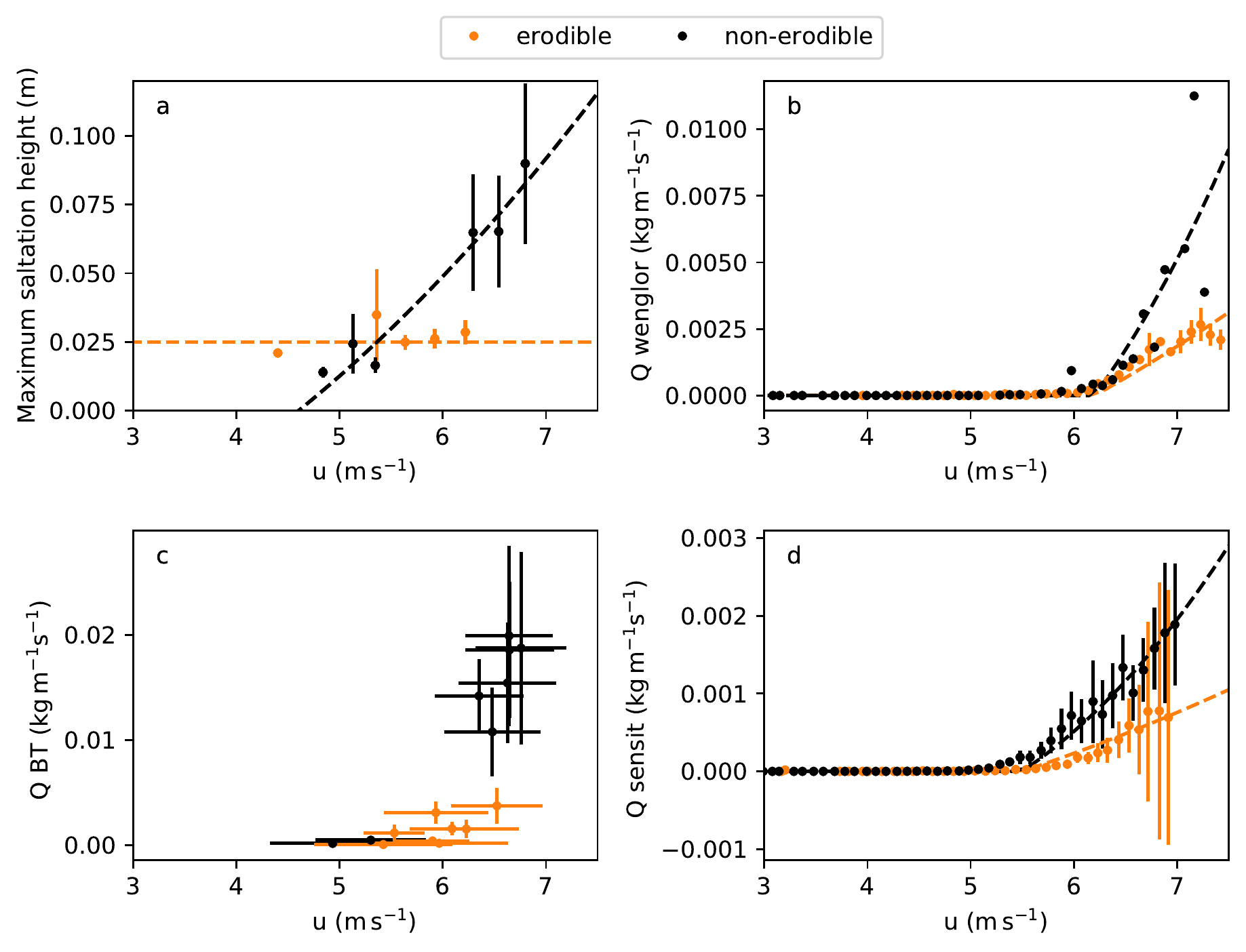}
\caption{Saltation height (a) and sediment flux (Q) as a function of wind velocity on both surfaces, as measured from Wenglor vertical array (b), Bagnold trap (c), and Sensit counters (d).\label{Fig2} }
\end{figure}

Our measurements show evidence of different particle behavior over the erodible and non-erodible beds. We find that the saltation height on the erodible surface is invariant with wind velocity whereas it increases with wind velocity on the non-erodible surface, as has been noted by other researchers \cite[Figure \ref{Fig2}a]{ bagnold1937transport, bagnold1941physics, creyssels2009saltating, ho2012particle, martin2017wind}. This field measured saltation height behavior then drives a change in sediment transport law on the erodible and non-erodible surface, as confirmed by our three independent measures of sand transport: a vertical array of Wenglor laser counters (Figure \ref{Fig2}b), Bagnold type sand traps (Figure \ref{Fig2}c), and Sensit piezoelectric counters (Figure \ref{Fig2}d).

Figures \ref{Fig2} b, c, and d show that for a given wind velocity, the amount of sand transported over the non-erodible surface is greater than that transported over the erodible surface. According to \citeA{bagnold1937transport}, the velocity of saltating grains over the erodible bed is independent of the wind velocity, and consequently the sand flux over an erodible surface scales quadratically with the wind speed \cite[ orange dashed lines Figure \ref{Fig2}b and d]{ungar1987steady, werner1990steady}. However, over the non-erodible bed, the particle velocity increases with wind velocity, thereby establishing a cubic dependence of sand transport on wind velocity \cite[ black dashed lines Figure \ref{Fig2}b and d]{ho2011scaling}. Two equations can thus be proposed to fit our datasets: 
\begin{equation}
\label{Eq1}
    Q_{\mathrm{sat}} = p\, Q_{\mathrm{ref}} \frac{u^2 -u_t^2 }{u_t^2},
\end{equation}
for the erodible surface datasets, and, 
\begin{equation}
\label{Eq2}
    Q_{\mathrm{sat}} = p\, Q_{\mathrm{ref}} \frac{u^2 -u_t^2 }{u_t^2} \frac{u}{u_t}, 
\end{equation}

for the non-erodible surface datasets, with $Q_{\mathrm{ref}}$ as the reference flux that is dependent on the sand characteristics, $u_t$, the threshold velocity, and $p$, a fitting parameter (see Supplementary Information for details on values for each measurement method). Because of this change in transport law, to respect mass balance, the transition from non-erodible to erodible bed should thus generate sand deposition. 

\section{Bedform Development}

\subsection{Conceptual Model}

\begin{figure}
\includegraphics[width=\textwidth]{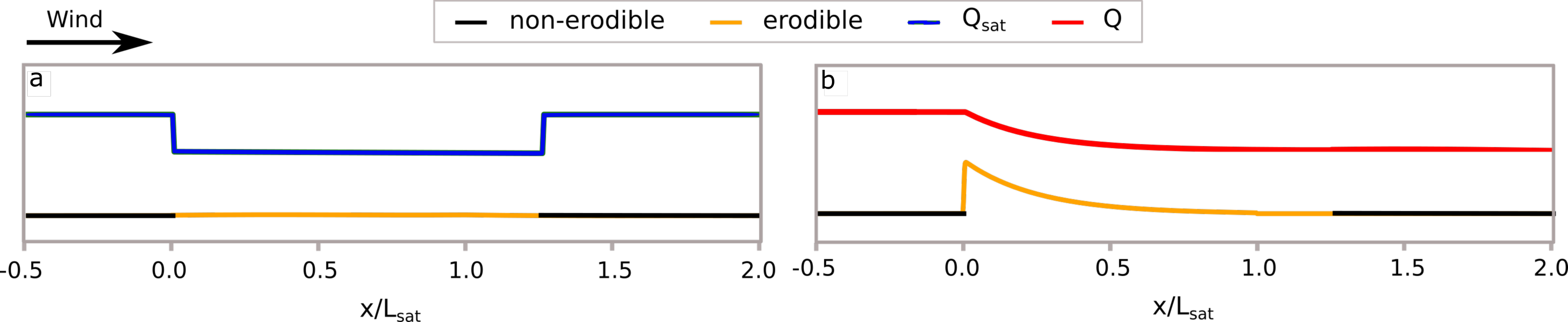}
\caption{Conceptual model for emergence of a sand patch driven by change in sand transport in the case of limited sand availability surface. (a) Pre-deposition state with the associated potential saturated sand flux (blue line). (b) Post-deposition state, with red line representing the actual sand flux. \label{Fig3}}
\end{figure}

Based on our field measurements, we propose a conceptual model to explain the emergence of an isolated sand patch on a flat, non-erodible bed with limited sand availability. We consider a flat, non-erodible surface (represented in black on Figure \ref{Fig3}a) adjacent to an erodible zone (in orange). Due to this change in surface characteristics, and according to equations \ref{Eq1} and \ref{Eq2}, a drop in the saturated sand flux at the transition from the non-erodible to erodible surface should occur (blue line on Figure \ref{Fig3}a). However, the flux does not adjust instantaneously to its new saturated value, but responds with a characteristic relaxation length, called the saturation length $L_{\mathrm{sat}}$, to reach $Q_{\mathrm{sat}}$ \cite{sauermann2001continuum, andreotti2010measurements, pahtz2013flux, selmani2018aeolian}. The red line represents this decrease in sand flux downwind of the non-erodible/erodible bed boundary (Figure \ref{Fig3}b). To respect mass balance, the excess sand transported on the non-erodible surface must deposit at the non-erodible/erodible transition following the decrease in sand flux over $L_{\mathrm{sat}}$, which thereby leads to the formation of a sand deposit (Figure \ref{Fig3}b). The rapid decrease in sand flux at the transition from a non-erodible to erodible surface (red line) thus generates a sand patch with an asymmetric shape, possessing a sharp upwind edge with a smooth downwind tail (Figure \ref{Fig3}b).

This simple conceptual model assumes a constant wind velocity above threshold, and a sharp transition from a non-erodible to erodible surface. In the next section, we compare qualitatively the topography of an incipient bedform in the field to the idealized patch presented in Figure \ref{Fig3}b.

\subsection{Field Evidence}

Sand transport measurements over a centimeter-high initial sand patch are challenging in the field as the placement of instruments can modify or destroy the emerging bedform by disrupting the windflow. Consequently, we measure concurrently the topography of an emerging sand patch and the saltation layer height with a non-invasive TLS. According to the measurements presented in Figure \ref{Fig2}a, we can use the dependence of the saltation layer height upon the wind velocity as a proxy for the appropriate transport law. To confirm that the change in sand flux acts as a driver for sand patch initiation, we measured the topography and saltation layer height pre-(black) and post-(orange) emergence of a sand patch on a sediment availability-limited, non-erodible surface (Figure \ref{Fig4}; field site and method are described in Supplementary Information). Figure \ref{Fig4} shows the height of the saltation layer is constant above the non-erodible surface, whereas it decreases over the developing patch due to its erodible sand surface. When sand particles start to travel over the erodible surface, each grain impact with the bed generates a particle ejection (splash effect), so that this process \remove{is} consumes energy. Consequently, saltating particles lose energy and experience a lower jump height, causing a decrease in the height of the saltation layer \cite{bagnold1937transport, ho2012particle, ho2014aeolian, valance2015physics}.

\begin{figure}
\includegraphics[width=\textwidth]{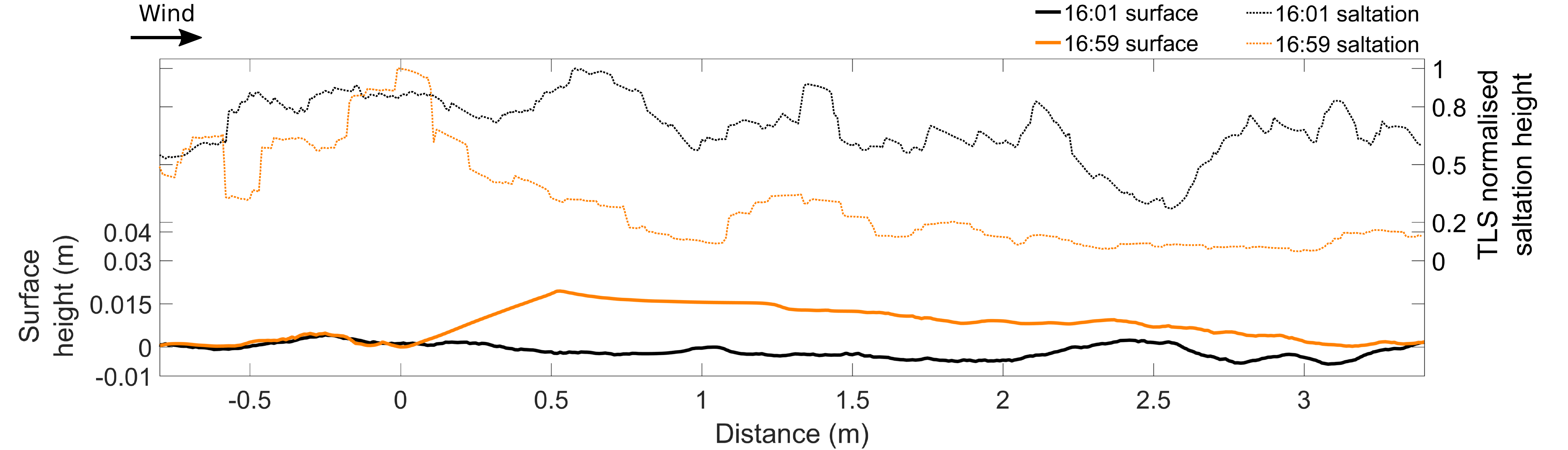}
\caption{TLS measured surface over an hour during the initial development of a sand patch and the corresponding relative saltation height over the same surface. Measurements were undertaken in the Great Sand Dunes National Park. The average wind speed measured at 0.1m above the surface during the experiment was 6.35 m$\,s^{-1}$. Relative saltation height is normalized by the maximum saltation height within each x-minute measurement period (the methods are detailed in the Supplementary Information).\label{Fig4}}
\end{figure}

As predicted by our conceptual model (Figure \ref{Fig3}), the observed initial sand patch exhibits a reverse asymmetry, with the steepest slope at the upwind edge. Our field measurements (Figure \ref{Fig4}) show a rapid decrease in saltation height from the upwind edge of the patch to a distance 1.4$\pm$0.3 m downwind of the patch toe. According to our conceptual model, sand deposition occurs over the saturation length. Although the relationship between $L_{\mathrm{sat}}$ and the grain diameter is still a matter of debate \cite{pahtz2013flux, pahtz2017fluid, selmani2018aeolian}, here we follow \citeA{andreotti2010measurements} to estimate $L_{\mathrm{sat}}$ as
\begin{equation}
    L_{\mathrm{sat}} \approx 2.2\, \frac{\rho_s}{\rho}\, d
    \label{Eq3}
\end{equation}

At the Great Sand Dunes field site, the grain size is d=350 $\pm$ 50 $\mu$m, mass density is $\rho_s$ =2650 kg$\,$m$^{-3}$, and the air density $\rho$ = 1.2 kg$\,$m$^{-3}$ that yields a saturation length of 1.7 $\pm$ 0.25 m, in good agreement with our field measurements. This therefore suggests that the saturation length sets the length of the incipient sand patch.

\section{Discussion and Conclusions}

Combining field measurements and a simple physically-based model, we propose a mechanism to explain the initiation of aeolian sand patches where there is limited sand availability. A change in surface characteristics (erodible/non-erodible or dry/moist) is critical, and leads to a modification of the sand transport dynamics. In agreement with previous studies, we show that the quantity of transported sand, and height of particle saltation, drops when encountering an erodible surface. The corresponding decrease in sand flux generates deposition in order to satisfy mass balance, thus adding sediment to the patch. Moreover, our field measurements demonstrate that the saturation length controls the size of the emerging deposit associated with the spatial relaxation of flux. Besides a change in surface mobility, the second critical parameter controlling sand patch emergence is the incoming sand flux. In our conceptual model, we assume the incoming sand flux equals the saturated sand flux associated with the non-erodible surface. However, the value of incoming flux depends largely on the sand source availability upwind of the initial patch. Without appropriate sand supply, such incipient bedforms are likely to degrade rapidly \cite{lancaster1996field, nield2011surface}. The majority of sand patches develop in interdune areas \cite{lancaster1996field} and beaches \cite{hesp1997crescentic, nield2011aeolian, baddock2018early, hage2018determining, montreuil2020early}, and in these cases sand sources are provided by the surrounding dry sandy surfaces. However, in the case of a succession or field of patches, if all the excess sand is deposited on the upwind erodible surfaces (as in the case of our conceptual schematics), then sediment supply would be further reduced to downwind patches. This condition likely creates a control on sand feeding of downwind patches and suggests there is a role for temporal wind fluctuations, both in strength and direction, in maintaining a broad field of multiple sand patches.
 As sand starts to be deposited, the initial bedform will interact with the wind flow and consequently the downwind variation of the sand flux will depend not only on the nature of the substrate (erodible/ non-erodible) but also on the underlying and developing topography \cite{claudin2013field, courrech2015dune, bristow2022topographic}. Consequently, to develop the conceptual arguments presented herein and investigate the conditions under which the aeolian sand patch is most likely to evolve, the present model needs further development to include full coupling between wind, transport and topography. In order to examine propagative solutions in a simplified dune model that accounted for these couplings, \citeA{andreotti2002bselection} identified flat bedform profiles without slipfaces (patches), but these solutions did not account for the change of transport law when bed conditions varied. However, these results did show the necessity of an incoming flux for these solutions to exist. The present study shows, for the first time, that it is possible to develop a sand patch on a non-erodible surface without any additional perturbation from the topography of the bed, and opens the way for study of the evolution of isolated sand patches towards larger bedforms and fully developed dunes \cite{kocurek1992dune, bristow2022topographic}.

\section*{Data Availability}
The data used in this manuscript can be found in the NERC National Geological Data Center:  Huab river valley dataset \cite<https://doi.org/10.5285/99e4446f-c43a-492d-83c9-e896206649c0,>{huabdata} and Great Sand Dunes National Park dataset  \cite<https://doi.org/10.5285/46e9ff95-27ca-4d3b-b587-fc9ce22c5781,>{medanodata}. Supplementary figures and text can be found in the supporting information.
%


\acknowledgments
This work was funded by the TOAD (The Origin of Aeolian Dunes) project (funded by the Natural Environment Research Council, UK and National Science Foundation, USA;
NE/R010196NSFGEO‐NERC, NSF-GEO‐1829541 and NSF‐GEO‐1829513). Research was undertaken at GSD under a Scientific Research and Collection permit GRSA‐2018‐SCI‐004, and we are very grateful for support from A. Valdez and F. Bunch. For the Huab fieldwork, we acknowledge Gobabeb Namib Research Institute, J. Kazeurua, I. Matheus, L. Uahengo, MET and NCRST (permits 1913/2014; 2051/2015; 2168/2016, RPIV00022018). Data processing used the IRIDIS Southampton Computing Facility. J. M. Nield was supported by a Department of Geology and Geophysics, Texas A$\&$M University, Michel T Halbouty Visiting Chair during the GSD field campaign. We thank B. Andreotti, C. Gadal, C. Narteau and TOAD project partners for useful discussions. We
also thank Patrick Hesp and an anonymous reviewer for their careful reading of our
manuscript and their insightful comments and suggestions.
\\
\clearpage

\title{Supporting Information for "Field evidence for the initiation of isolated aeolian sand patches"}
%
%

%
%



\authors{P. Delome \affil{1,2}, J. M. Nield \affil{1}, G. F. S. Wiggs\affil{3}, M. C. Baddock\affil{4}, N. R. Bristow\affil{5}, J. L. Best \affil{6}, K. T. Christensen \affil{7}, and P. Claudin\affil{8}  }


\affiliation{1}{School of Geography and Environmental Science, University of Southampton, Southampton, UK}
\affiliation{2}{now at: Energy and Environment Institute, University of Hull, Hull, UK}
\affiliation{3}{School of Geography and the Environment, University of Oxford, Oxford, UK}
\affiliation{4}{Geography and Environment, Loughborough University, Loughborough, UK }
\affiliation{5}{Mechanical Engineering, St Anthony Falls Laboratory, University of Minnesota, Minneapolis, USA}
\affiliation{6}{Departments of Geology, Geography and GIS, Mechanical Science and Engineering and Ven Te Chow Hydrosystems Laboratory, University of Illinois at Urbana-Champaign, USA}
\affiliation{7}{Departments of Mechanical, Materials and Aerospace Engineering and Civil, Architectural and Environmental Engineering, Illinois Institute of Technology, USA}
\affiliation{8}{Physique et Mécanique des Milieux Hétérogènes, CNRS, ESPCI Paris, PSL Research University, Université Paris Cité, Sorbonne Université, Paris, France}

%
%

%

%
%

\noindent\textbf{Contents of this file}
\begin{enumerate}
\item Text S1 to S4
\item Figures S1 to S3
\end{enumerate}

\section{Introduction}
Our study involved two field campaigns. Data collected in the Skeleton Coast National Park, Namibia, was used to confirm the differences in sand transport dynamics on erodible and non-
erodible surfaces. The saltation dynamics that occurred during the emergence of a sand patch on a non-erodible surface were measured at Great Sand Dunes National Park, Colorado, USA. The following section contain the description of the field sites and the methods used to collect and analyze the data.


\section{Text S1-Field sites}
The study area in the Huab Valley dune field of the Skeleton Coast National Park, northwest Namibia, consisted of a flat surface covered with gravel and an ephemeral sand patch aligned with the predominant south-southwest winds \cite{lancaster1982dunes}. The gravel acts as an armor layer, which makes the surface non-erodible, in contrast to sand patches that are erodible. We measured sand transport, saltation height and wind velocity over both surfaces during periods of high sand transport. The measurements were performed simultaneously over both surfaces, a co-located sand patch and gravel surface from the 13$^{th}$ to the 15$^{th}$ September 2019.
The Medano Creek site, close to the Visitor Center at Great Sand Dunes National Park, Colorado, USA, consisted of a moist sand surface, adjacent to an ephemeral creek bed, with sand being
supplied upwind from a field of protodunes. The moisture was large enough to prevent sand being eroded from the surface (i.e. it was a non-erodible surface), but not moist enough for adhesion to occur. We measured saltation height and surface change over this surface during an hour-long period on the 15$^{th}$ April 2019.

\section{Text S2-Wind Velocity and Sand Transport (Skeleton Coast National Pak, Namibia)} 
\begin{figure}
\includegraphics[width=\textwidth]{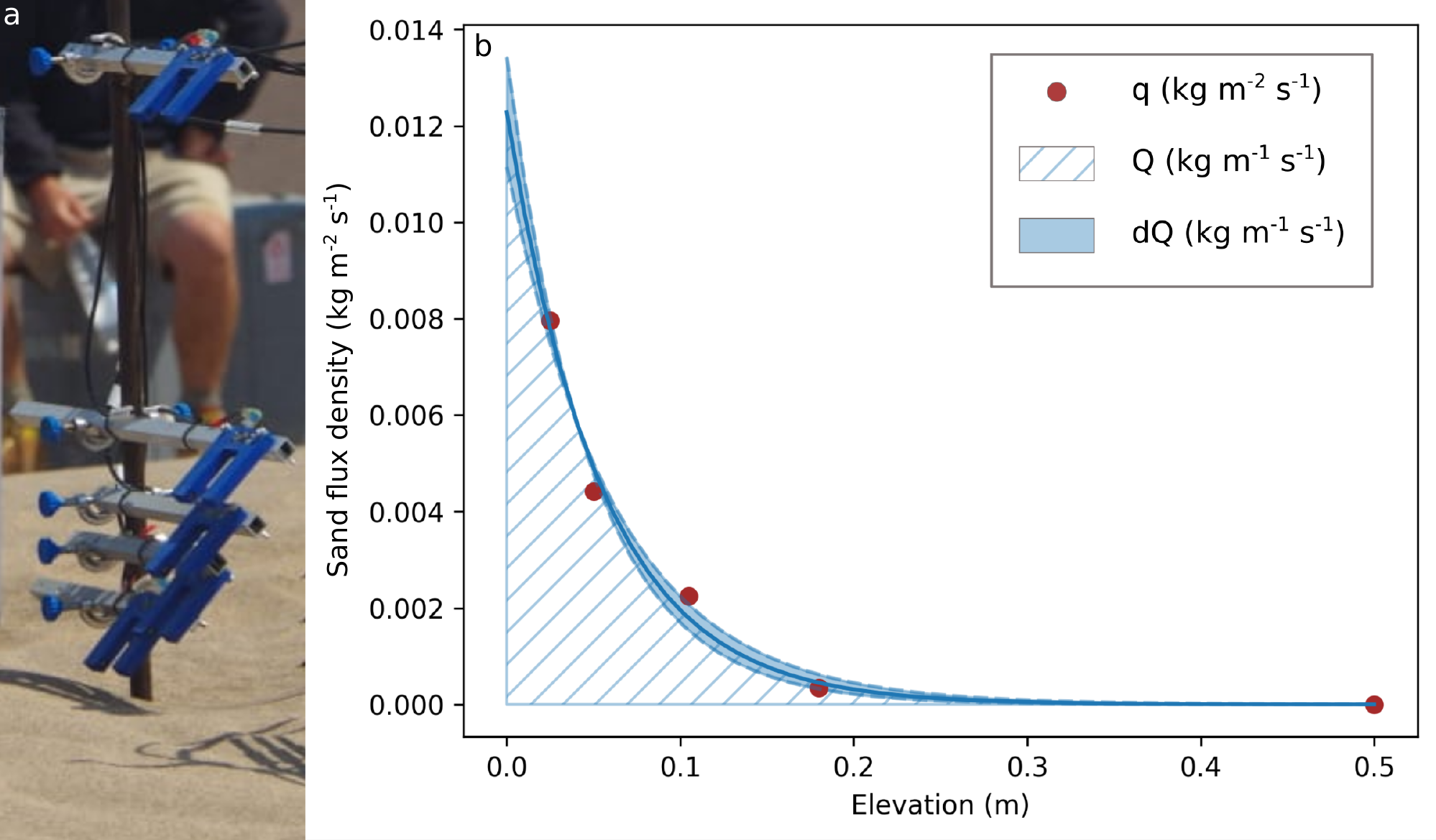}
\caption{a) Vertical array of Wenglor counters used in the field. b) Sand flux measured at each elevation (red dots), the hatched
area is the total flux Q, and the blue area is the error calculated from the exponential fit. \label{FigS1}}
\end{figure}

We measured the wind velocity at a height of 0.085 m with a hotwire anemometer (DANTEC 54T35 probes), at a frequency of 0.1 Hz. We used three different methods to quantify sand transport on both erodible and non-erodible surfaces. i) Five laser particle counters (Wenglor
YH03PCT8) were positioned in a vertical array at height of 0.025, 0.05, 0.115, 0.18 and, 0.5 m measuring at a frequency of 0.1 Hz (Figure \ref{FigS1}a).
ii) A Sensit contact particle counter, which counts sand particles within 0.03 m from the surface recorded total counts every 10 seconds. iii) A modified Bagnold sand trap sampled moving sand between 0 to 0.5 m above the surface over a time period varying between 10 and 30 minutes (Figure \ref{FigS2}a). The sand traps allowed us to quantify the physical characteristics of the transported sand. We measured the grain size using a laser granulometer (Malvern Mastersizer 3000), and the density of the sand using a pycnometer. We then used this particle size distribution to convert particle counts from the Wenglors into sand flux following the method of \citeA{barchyn2014particle}. All the datasets are
subdivided into 30-sec intervals. The methods used to estimate the sand flux are described
below:
\begin{itemize}
    \item Wenglor array:
    The particle count recorded by each Wenglor instrument is converted into sand flux using the method proposed by \citeA{barchyn2014particle}. For each time period, we integrated the sand flux density at each elevation, $q$, using the corresponding exponential law to obtain the sand flux, $Q$
(equation \ref{EqS1} and Figure \ref{FigS1}b).
\begin{equation}
    Q = \int_{\mathrm{z=0}}^{\mathrm{z= \infty}} q_0 exp\left(\frac{-z}{z_0}\right) dz,
    \label{EqS1}
\end{equation}
where $q_0$ and $z_0$ are the parameters of the exponential fit.

\item Bagnold sand traps:
We positioned four Bagnold traps approximately 50 cm from each other and spanwise to incoming sand transport. Each trap comprised 25 slots of 2 cm-height. By measuring the amount of sand trapped in each slot, we could estimate the sand flux density at each elevation, $q$. We finally fit the exponential law (equation \ref{EqS1}) to the averaged sand flux (averaged over the four Bagnold traps), to calculate the sand flux, $Q$ (Figure \ref{FigS2}).

\begin{figure}
\includegraphics[width=\textwidth]{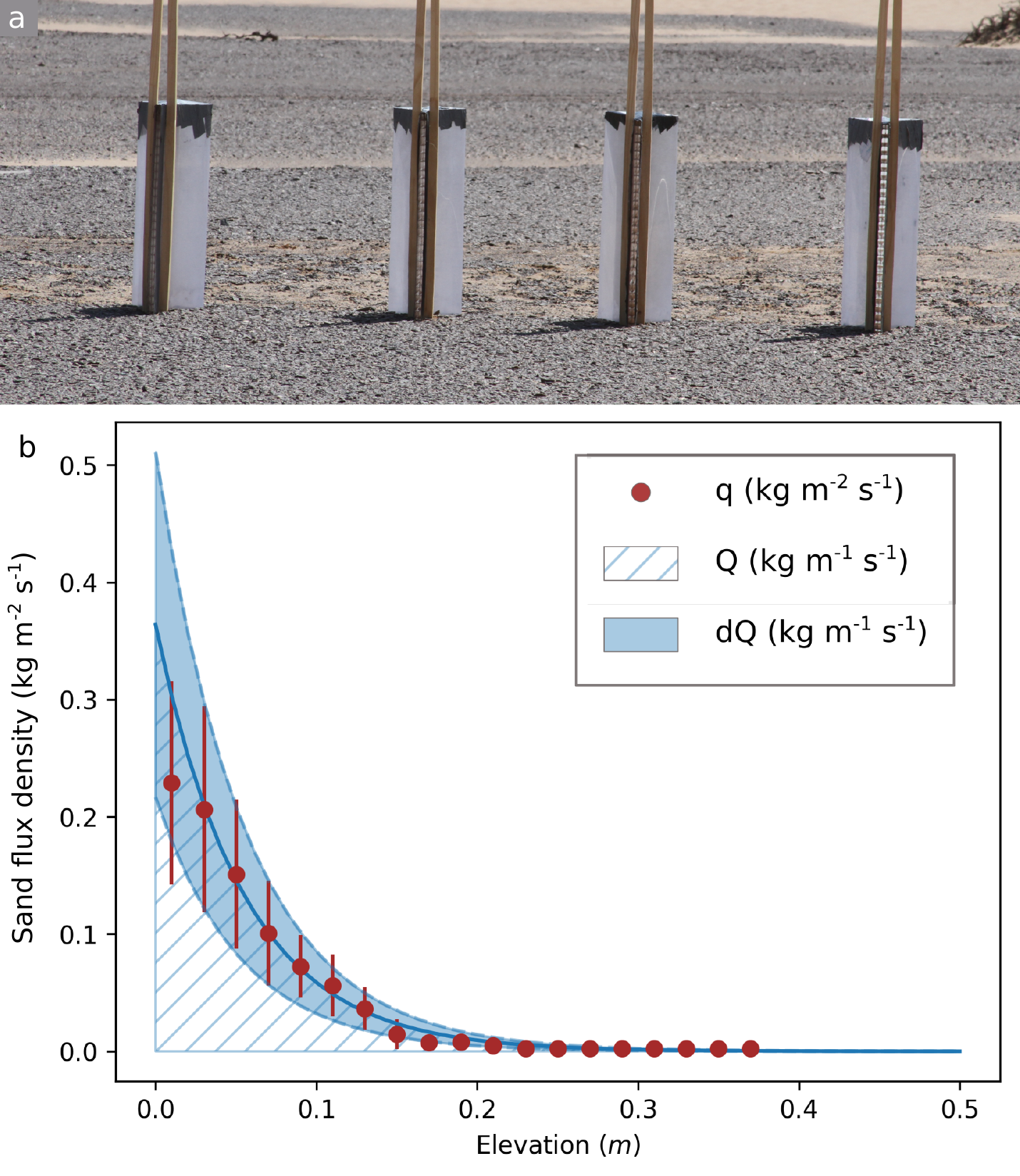}
\caption{a) Frontal view of Bagnold traps in the field. b) Sand flux measured at each elevation (red dots), the hatched area is the
total flux $Q$, and the blue area is the error calculated from the exponential fit. \label{FigS2}}
\end{figure}

\item Sensit-piezoelectric counter:
The Sensit piezoelectric counter detects particle impacts from all directions between 0 and 3 cm above the surface. We converted the particle counts into a flux using:
\begin{equation}
    Q= \frac{nV\rho}{A} h, 
\end{equation}
where $n$ is the counts per second, $V$ the volume of each sand particle, $A$ the area of the Sensit and $h$ the measurement height.
\end{itemize}

To compare transport at similar wind speeds on different surfaces, we concatenated all the data to generate datasets for erodible and non-erodible surfaces that we then binned according to the wind velocity, with an increment of 0.1 m$\,$s$^{-1}$. Finally, we fitted the transport law, Equation (1) and (2), to the two datasets, erodible and non-erodible, respectively. The reference
flux $Q_{\mathrm{ref}}$ is defined as,

\begin{equation}
    Q_{\mathrm{ref}} = \rho_s d \sqrt{\frac{\rho_s}{\rho} g d},
\end{equation}

equals 1.8 kg$\,$m$^{-1}$s$^{-1}$. Using the Wenglor counter dataset, we find a threshold velocity $u_t$ = 6.2 m$\,$s$^{-1}$ (at 0.085 m), and the fitting parameter $p$ equals 0.004 and 0.009 respectively for the erodible and non-erodible surfaces. For the Sensit datasets, we obtain $u_t$ = 5.5 m$\,$s$^{-1}$ and $p$ equals 0.0007,
and 0.0014 respectively for the erodible and non-erodible surfaces. Due to the intermittency of the measurement with the Bagnold traps, there was insufficient data to fit a transport law. However, the Bagnold trap data show qualitatively the difference in transport capacity over each surface.

\section{Text S3-Maximum Saltation Height and Surface Change (Skeleton Coast National Park and Great Sand Dunes National Park)} 

We measured the maximum saltation height over each surface using terrestrial laser scanning (TLS), following the methods of \citeA{nield2011application}. An area of approximately 1 m$^2$ was
scanned using a Leica P20 TLS over a 3-minute period, and a filter of 35$^{°}$ was applied to separate laser returns from above, and on, the surface. Data was gridded in 0.01 m$^2$ to minimize the impact
of mixed pixels or large-scale topography. Within each scan, saltation heights were obtained for each grid square based on the maximum height of the laser returns above the surface and the minimum height of the surface. The maximum value for each scan was assigned a mean velocity from the hotwire anemometer. Data were then binned for each surface using 0.3 m$\,$s$^{-1}$ velocity brackets, with a minimum of four points in each bracket and points where the standard deviation was greater than the mean value within the bracket were excluded.

Additionally, at Great Sand Dunes National Park we used two TLS instruments, a Leica P20 and a Leica P50. As the sand patch that developed was the same order of magnitude in height as the saltating grains, we first concatenated four consecutive scans of the surface and applied the saltation filter to maximize the chance that scanning of the surface was not occluded by saltation. We then use each individual scan to plot the surface (gridded at 0.0001 m$^2$) and saltation (gridded at 0.01 m$^2$) at the start and end of the measurement period (1 hour).

\section{Text S4-Estimation of the Saturation Length from the Evolution of the Saltation Layer Height}

\begin{figure}
\includegraphics[width=\textwidth]{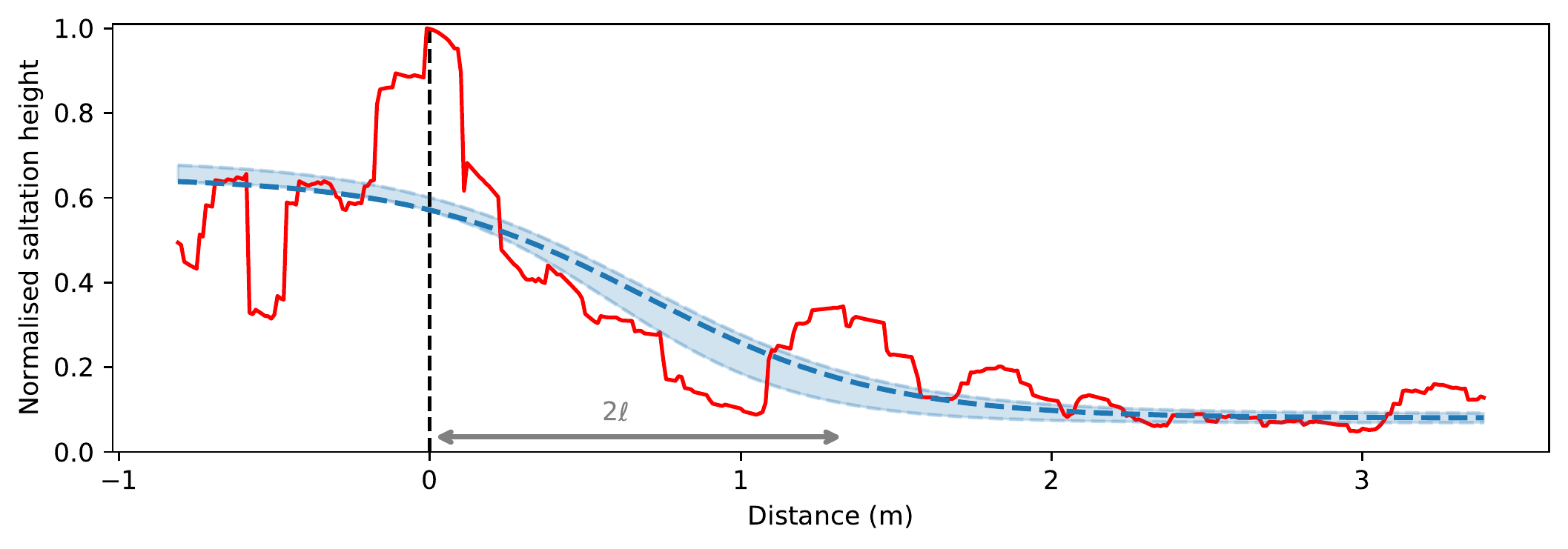}
\caption{Evolution of the saltation layer height (red line) with distance from the sand patch upwind toe (see orange line on Figure
4). The black vertical dashed marks the transition from non-erodible to erodible substrate. The dashed blue line is the sigmoidal
fit, with the uncertainty represented by the blue shading. \label{FigS3}}
\end{figure}

The transition appears as a decreasing sigmoidal curve (Figures 4 and \ref{FigS3}). To quantify this observation, we fit a hyperbolic tangent to the height profile, defined as,

\begin{equation}
    H = \left(H_{ne} - H_{e}\right) \frac{tanh (-(x-x_0)/l + 1}{2} + H_e,
    \label{EqS3}
\end{equation}

where $H_e$ and $H_{ne}$ are the normalized saltation height over the erodible and non-erodible surface, $x_0$ is the location of the transition and $l$ is the exponential length.

We find that the normalized saltation height plateaus at about 0.67 on the non-erodible surface ($H_{ne}$), and at about 0.08 on the erodible surface ($H_e$). The transition between these two areas is a smooth transition with a characteristic length 2$l$ = 1.4$\pm$ 0.3 m.

%
%


%
%
%
%
%


%
\bibliography{Bibliography_TL} 
%




%
%
%
%
%

\end{document}